\documentstyle[preprint,aps]{revtex}
\newcommand{\be}{\begin{equation}}
\newcommand{\ee}{\end{equation}}
\begin{document}
\draft
\title{Seesaw Mechanism in Three Flavors}
\author{
	Haijun Pan$^1$\thanks{Email: phj@mail.ustc.edu.cn}, and 
	G. Cheng$^{2,3}$\thanks{Email: gcheng@ustc.edu.cn}}
\address{$^1$Lab of Quantum Communication and Quantum Computation, 
	and Center of Nonlinear Science, \\
	University of Science and Technology of China,
	Hefei, Anhui, 230026, P.R.China}
\address{$^2$CCAST(World Laboratory), P.O.Box 8730, Beijing 10080,
	P.R.China}
\address{$^3$Department of Astronomy and Applied Physics, \\
	University of Science and Technology of China, 
	Hefei, Anhui, 230026, P.R.China}
\maketitle

\begin{abstract}
We advance a method used to analyse the neutrino properties (masses and 
mixing) in the seesaw mechanism. Assuming quark-lepton symmetry and 
hierarchical light neutrino masses, we establish rather simple relations 
between the light and the heavy neutrino parameters in the favored regions 
of the solar and the atmospheric neutrino experiments. A empirical condition 
satisfied by the right-handed mixing angles is obtained.
\end{abstract}

\section{Introduction}

Whether neutrinos have nonzero masses or not? How large would the mixing 
angles be? Are they like that in the quark sector? Those are among the 
pressing questions in particle physics. The solar \cite{solar} and 
atmospheric \cite{atm} neutrino data suggest that neutrinos do have 
masses and the recent results from Super-Kamiokande (SK) \cite{atm} 
imply a nearly maximal mixing of $\nu_{\mu }$ and $\nu_{\tau }$. 
In another hand, the fact that neutrinoless double-$\beta $ decay and 
other lepton number nonconserving processes are not observed experimentally 
reflects the smallness of the neutrino masses \cite{bilenky}. The seesaw 
mechanism \cite{seesaw} has a natural explanation for the small neutrino 
masses and may enhance lepton mixing up to maximal \cite{smirnov,falcone,list}.

According to the seesaw mechanism, at $M\gg m_D$, the Majorana mass matrix 
$m^{{\rm eff}}$ of the left-handed (LH) neutrino components is given as \cite{smirnov} 
\be
m^{{\rm eff}}=m_DM^{-1}m_D^T. 
\ee
Here $M$ is the Majorana mass matrix of the right-handed 
(RH) neutrino components and $m_D$ is the neutrino Dirac mass matrix 
which could be equal to the mass matrix of the up quarks: $m_D=m^{up}$ 
according to some kind of quark-lepton symmetry \cite{smirnov,falcone,kuo1}. 
In the basis where $M^{-1}$ is diagonal, 
$M^{-1}=M_i^{-1}\delta_{ij}\equiv R_i^2\delta_{ij}~(i,j=1,2,3)$  
$m_D$ can be written as \cite{kuo1}  
\be
m_D=U_0m_D^{{\rm diag}}V_0.
\ee
Here $U_0$ and $V_0$ are LH and RH rotations respectively and 
$m_D^{\rm diag}={\rm diag}\left\{ m_1,m_2,m_3\right\} $.

In this paper, we study a problem what we can know about the masses and 
mixing of the right handed neutrinos from the low energy neutrino data. 
The paper is organized as follows. In Sec. II a parameterization is 
introduced and the seesaw mechanism is expressed in two formula: one of 
them involves only the neutrino masses and the other involves only some 
nondimensional parameters such as mass ratios and mixing angles. Then 
the RH neutrino masses and mixing angles are derived. In Sec. III we get 
rather simple relations between the masses and mixing angles entering 
the seesaw formula in the favored regions of the solar and atmospheric 
experiments. The numerical results they infer are given whereafter. 
We summarize and discuss our main results in Sec. IV.


\section{General framework}

\subsection{Parameterization}

Since the CP-violating effects in neutrino oscillations should be small 
\cite{cp}, we shall therefore ignore it and so $U_0$ and $V_0$ are real 
orthogonal matrices. For simplicity, We also set $U_0\sim I$. That is, 
the left-handed rotations that diagonalize the charged lepton $m_1$ and 
neutrino Dirac mass matrices $m_D$ are the same or nearly the same and 
so the large lepton mixing results from the seesaw transformation \cite{smirnov}. 
Under these assumptions, it is convenient to write 
\be
m_D^{{\rm diag}}V_0M^{-1}V_0^Tm_D^{{\rm diag}}
=U_0^TU\left(N^{{\rm diag}}\right)^2U^TU_0
\approx U\left(N^{{\rm diag}}\right)^2U^T, 
\ee
or by inverting it,
\be
\left(m_D^{{\rm diag}}\right)^{-1}U\left(N^{{\rm diag}}\right)^2U^T
\left(m_D^{{\rm diag}}\right)^{-1}=V_0M^{-1}V_0^T 
\ee
where $U$ is LH rotation induced by $M^{-\frac{1}{2}}$ and 
$N^{\rm diag}=diag\left\{ n_1,n_2,n_3\right\} $ 
with $n_i^2=m_i^{{\rm eff}}~\left(i=1,2,3\right)$, 
the eigenvalues of $m^{{\rm eff}}$.

Analogy with that in the two flavors case \cite{kuo2}, we introduce the 
following mass parameterization, 
\begin{mathletters}
\begin{eqnarray}
\xi_3 &=&\frac{1}{2}\ln \frac{m_2}{m_1},~~~~~
\xi_8=\frac{1}{6}\ln \frac{m_3^2}{m_1m_2}; \\
\eta_3 &=&\frac{1}{2}\ln \frac{R_1}{R_2},~~~~~
\eta_8=\frac{1}{6}\ln \frac{R_1R_2}{R_3^2}; \\
\kappa_3 &=&\frac{1}{2}\ln \frac{n_2}{n_1},~~~~~	
\kappa_8=\frac{1}{6}\ln \frac{n_3^2}{n_1n_2}
\end{eqnarray}
\end{mathletters}
and the mixing matrices are parameterized as usual, 
\begin{mathletters}
\begin{eqnarray}
U &=&	\exp \left(i\theta_{23}\lambda_7\right)
	\exp \left(i\theta_{13}\lambda_5\right)
	\exp \left(i\theta_{12}\lambda_2\right), \\
V_0 &=&	\exp \left(i\beta_{23}\lambda_7\right)
		\exp \left(i\beta_{13}\lambda_5\right)
		\exp \left(i\beta_{12}\lambda_2\right).
\end{eqnarray}
\end{mathletters}
Here, $\lambda_2$, $\lambda_5$, $\lambda_7$ are Gell-mann matrix. 
One can see that $\eta_3$ and $\eta_8$ describe the hierarchy of the 
RH neutrino masses and are always nonnegative. Especially, $\eta_3=0$ 
implies $M_1=M_2$ while $\eta_3=3\eta_8$ implies $M_2=M_3$. 
Using the diagonal Gell-mann matrix $\lambda_3$ and $\lambda_8$, 
the mass matrices involving now can be rewritten as
\begin{mathletters}
\begin{eqnarray}
m_D^{{\rm diag}} &=& \left(m_1m_2m_3\right)^{\frac{1}{3}}
   			   e^{-\xi_3\lambda_3-\sqrt{3}\xi_8\lambda_8}, \\
M^{-1} &=& \left(R_1^2R_2^2R_3^2\right)^{\frac{1}{3}}
	     e^{2\eta_3\lambda_3+2\sqrt{3}\eta_8\lambda_8}, \\
\left(N^{{\rm diag}}\right)^2 &=& 
		\left(n_1^2n_2^2n_3^2\right)^{\frac{1}{3}}
		e^{-2\kappa_3\lambda_3-2\sqrt{3}\kappa_8\lambda_8}.
\end{eqnarray}
\end{mathletters}

This parameterization shows clearly that the relevant variables in the 
diagonalization of $M^{-1}$ are $\theta_{12}$, $\theta_{13}$, $\theta_{23}$, 
$\kappa_3$, $\kappa_8$, $\xi_3$ and $\xi_8$. Of these, it is usually 
assumed that $\xi_3$ and $\xi_8$ can be identified with the corresponding 
quantities of the up sector of quarks as stated before and $\theta_{12}$, 
$\theta_{13}$, $\theta_{23}$, $\kappa_3$, $\kappa_8$ can be obtained, at least 
approximately, from the low energy neutrino data. Now let us denote 
\begin{eqnarray}
\label{ovX} 
\overline{X}\left(\kappa,\xi,\theta\right) 
	&=& 	\left(R_1^2R_2^2R_3^2\right)^{\frac{1}{3}}
		V_0e^{2\eta_3\lambda_3+2\sqrt{3}\eta_8\lambda_8}V_0^T \\
	&=&	\left(m_1m_2m_3\right)^{-\frac{2}{3}}
		\left(n_1^2n_2^2n_3^2\right)^{\frac{1}{3}}
		X\left(\kappa,\xi,\theta\right).  \nonumber
\end{eqnarray}
Here
\be
\label{X}
X\left(\kappa,\xi,\theta\right)
=e^{\xi_3\lambda_3+\sqrt{3}\xi_8\lambda_8}U
 e^{-2\kappa_3\lambda_3-2\sqrt{3}\kappa_8\lambda_8}U^T
 e^{\xi_3\lambda_3+\sqrt{3}\xi_8\lambda_8}  
\ee
and $\kappa$, $\xi$ and $\theta$ refer to $\kappa_3$, $\kappa_8$; $\xi_3$, 
$\xi_8$ and $\theta_{12}$, $\theta_{13}$, $\theta_{23}$ respectively. 
Eq.\ (\ref{ovX}) is equivalent with the following two equations:
\begin{mathletters}
\begin{eqnarray}
\label{seesaw:1}
R_1^2R_2^2R_3^2&=&\left(m_1m_2m_3\right)^{-2}\left(n_1^2n_2^2n_3^2\right),\\
\label{seesaw:2}
X\left(\kappa ,\xi ,\theta \right)
&=&V_0e^{2\eta_3\lambda_3+2\sqrt{3}\eta_8\lambda_8}V_0^T.  
\end{eqnarray}
\end{mathletters}
The first relation is just the equality of the determinations of both sides 
of Eq.\ (\ref{ovX}). Taking the total term 
$\left(R_1^2R_2^2R_3^2\right)^{\frac{1}{3}}
=\left(m_1m_2m_3\right)^{-\frac{2}{3}}
 \left(n_1^2n_2^2n_3^2\right)^{\frac{1}{3}}$ 
out from Eq.\ (\ref{ovX}) we get the second relation. For late use, we present 
here the expression of the inverse of $X\left(\kappa ,\xi ,\theta \right)$. 
It is easy to know from Eq.\ (\ref{X}) that 
\be
X^{-1}\left(\kappa ,\xi ,\theta \right)
=e^{-\xi_3\lambda_3-\sqrt{3}\xi_8\lambda_8}U
 e^{2\kappa_3\lambda_3+2\sqrt{3}\kappa_8\lambda_8}U^T
 e^{-\xi_3\lambda_3-\sqrt{3}\xi_8\lambda_8}. 
\ee
So that 
\be
X^{-1}\left(\kappa ,\xi ,\theta \right)=X\left(-\kappa ,-\xi ,\theta \right)
\equiv Y\left(\kappa ,\xi ,\theta \right).  
\ee
and we have
\be
\label{Y}
Y\left(\kappa ,\xi ,\theta \right)
=V_0e^{-2\eta_3\lambda_3-2\sqrt{3}\eta_8\lambda_8}V_0^T.  
\ee
We will start from Eqs.\ (\ref{seesaw:2},\ref{Y}) to derive the expressions 
of $\eta_3$, $\eta_8$ and $V_0$. Then from Eq.~(\ref{seesaw:1}) 
$M_i\left(i=1,2,3\right)$ can be obtained. In following discussion, we shall 
omit the variables $\kappa ,\xi ,\theta $ in $X$ and $Y$.

\subsection{Determination of the Majorana masses}

In this subsection we deduce two equations about the hierarchy, $\eta_3$ and $\eta_8$, 
of the RH neutrino masses. Taking the trace of both sides of Eq.\ (\ref{seesaw:2}) 
we obtain 
\be
{\rm Tr}\left(V_0e^{2\eta_3\lambda_3+2\sqrt{3}\eta_8\lambda_8}V_0^T\right)
={\rm Tr}e^{2\eta_3\lambda_3+2\sqrt{3}\eta_8\lambda_8}={\rm Tr}X,  
\ee
that is,
\be
\label{A:1}
e^{2\eta_3+2\eta_8}+e^{-2\eta_3+2\eta_8}+e^{-4\eta_8}=X_{11}+X_{22}+X_{33}\equiv A.  
\ee
Similarly, taking the trace of both sides of Eq.\ (\ref{Y}) we get
\be
\label{B:1}
e^{-2\eta_3-2\eta_8}+e^{2\eta_3-2\eta_8}+e^{4\eta_8}=Y_{11}+Y_{22}+Y_{33}\equiv B.  
\ee

It is sufficient for solving $\eta_3$ and $\eta_8$ from Eqs.\ (\ref{A:1},\ref{B:1}) 
since $X_{ii}$ and $Y_{ii}$ ($i=1,2,3$) are known. Once $\eta_3$ and $\eta_8$ 
are solved, inserting $M_1=M_3e^{-2\eta_3-6\eta_8}$ , $M_2=M_3e^{2\eta_3-6\eta_8}$, 
$n_1^{{\rm 2}}=n_3^{{\rm 2}}e^{-2\kappa_3-6\kappa_8}$, and 
$n_2^{{\rm 2}}=n_3^{{\rm 2}}e^{2\kappa_3-6\kappa_8}$ in Eq.\ (\ref{seesaw:1}), 
we obtain the following expressions of the RH neutrino masses,
\be
\label{M:1}
M_1=Fe^{-2\eta_8-2\eta_3},~~~~~M_2=Fe^{-2\eta_8+2\eta_3},~~~~~M_3=Fe^{4\eta_8}.  
\ee
Here $F=\frac{m_t^2}{m_3^{{\rm eff}}}e^{4\kappa_8-4\xi_8}$ and we have 
identified $m_i\left(i=1,2,3\right)$ with the masses of up quarks.

All the above results are exact but formal. We need to decouple $\eta_3$ 
and $\eta_8$ in Eqs.\ (\ref{A:1},\ref{B:1}). From Eq.\ (\ref{A:1}), we have
\be
A=e^{2\eta_3+2\eta_8}+e^{-2\eta_3+2\eta_8}+e^{-4\eta_8}
\geq 3\left(e^{2\eta_3+2\eta_8}e^{-2\eta_3+2\eta_8}e^{-4\eta_8}\right)^{\frac{1}{3}}=3.
\ee
The equality is satisfied when $\eta_3=\eta_8=0$, that is, when $M_1=M_2=M_3$. 
At $A\gg 3$ (then $B\gg 3$ is also true), Eq.\ (\ref{A:1}) and Eq.\ (\ref{B:1}) 
can be approximated as follows,
\begin{mathletters}
\begin{eqnarray}
\label{A:2}
e^{2\eta_3+2\eta_8}+e^{-2\eta_3+2\eta_8}&\approx&A, \\
\label{B:2}
e^{2\eta_3-2\eta_8}+e^{4\eta_8}&\approx& B.  
\end{eqnarray}
\end{mathletters}
Such a case corresponds to at most two degenerate Majorana masses. 
There are now two possibilities to simplify the above two equations further:


{\em (a) }${\sl A>B}$

It is easy to know from Eqs.\ (\ref{A:2},\ref{B:2}) that $A>B$ 
implies $\eta_3>\eta_8$. So $e^{-2\eta_3+2\eta_8}$($<1$) may be omitted 
in Eq.\ (\ref{A:2}), 
\be
\label{A:3}
e^{2\eta_3+2\eta_8}=e^{2\eta_3-2\eta_8}e^{4\eta_8}\approx A,
\ee
Then it is easy to see from Eqs.\ (\ref{B:2},\ref{A:3}) that 
$e^{2\eta_3-2\eta_8}$ and $e^{4\eta_8}$ 
are roots of the following quadratic equation: 
\be
x^2-Bx+A=0. 
\ee
and the three eigenvalues of $X$ are
\begin{mathletters}
\label{A>B}
\begin{eqnarray}
e^{2\eta_3+2\eta_8} &\approx &A, \\
e^{-2\eta_3+2\eta_8} &\approx &\frac{2}{B-\sqrt{B^2-4A}},\\
e^{-4\eta_8} &\approx &\frac{2}{B+\sqrt{B^2-4A}}.
\end{eqnarray}
\end{mathletters}


{\em (b) }${\sl A<B}$

In this case, we have $\eta_3<\eta_8$. Omitting the term $e^{2\eta_3-2\eta_8}(<1)$ 
in Eq.\ (\ref{B:2}), we have
\be
e^{4\eta_8}=e^{2\eta_3+2\eta_8}e^{-2\eta_3+2\eta_8}\approx B. 
\ee
Now $e^{2\eta_3+2\eta_8}$ and $e^{-2\eta_3+2\eta_8}$ are roots of the 
following quadratic equation: 
\be
x^2-Ax+B=0. 
\ee
Thus one has  
\begin{mathletters}
\label{A<B}
\begin{eqnarray}
e^{2\eta_3+2\eta_8} &\approx &\frac{A+\sqrt{A^2-4B}}{2}, \\ 
e^{-2\eta_3+2\eta_8} &\approx &\frac{A-\sqrt{A^2-4B}}{2},\\
e^{-4\eta_8} &\approx &\frac{1}{B}.
\end{eqnarray}
\end{mathletters}

From Eq.\ (\ref{A>B}) we know that $e^{-2\eta_3+2\eta_8}\sim e^{-4\eta_8}$ 
(and so $M_2\sim M_3$) when $B^2\sim 4A$ and from Eq.\ (\ref{A<B}) 
$e^{2\eta_3+2\eta_8}\sim e^{-2\eta_3+2\eta_8}$ (and so $M_1\sim M_2$) 
when $A^2\sim 4B$. Far beyond these regions, both Eq.\ (\ref{A>B}) and 
Eq.\ (\ref{A<B}) give the same asymptotic solution: 
$e^{2\eta_3+2\eta_8}\approx A$, and $e^{-4\eta_8}\approx \frac{1}{B}$ 
and $e^{-2\eta_3+2\eta_8}=e^{-2\eta_3-2\eta_8}e^{4\eta_8}\approx \frac{B}{A}$. 
The solutions are also useful for rough estimation of the Majorana masses 
even when two of them are degenerate, which can be seen from 
$e^{2\eta_3+2\eta_8}<e^{2\eta_3+2\eta_8}+e^{-2\eta_3+2\eta_8}<2e^{2\eta_3+2\eta_8}$ 
and $e^{4\eta_8}<e^{2\eta_3-2\eta_8}+e^{4\eta_8}<2e^{4\eta_8}$. The maximal 
deviations for $e^{2\eta_3+2\eta_8} $ and $e^{4\eta_8}$\ are both $2$ times.

Usually one should have to solve a cubic characteristic equation to obtain 
the eigenvalues. In the seesaw model, however, one usually encounters such 
case where $e^{2\eta_3+2\eta_8}\gg 1$ and $e^{-4\eta_8}\ll 1$ simultaneously. 
This is a practical difficulty even in numerical calculation. More worse, 
the solution of a cubic equation is too ugly to see any relation between 
various physical quantities. By taking the trace of $X$\ and its inverse, 
we decompose the eigen-equation in two equations and each contains the main 
term of $e^{2\eta_3+2\eta_8}$ and $e^{4\eta_8}$ respectively. In concrete 
calculation, the expressions of $A$ and $B$ can be simplified to such a 
great extent that the dependence on the parameters can be exhibited explicitly. 
We will discuss this issue later.

\subsection{Determination of the RH angles}

Once one have the three eigenvalues solved, then the three eigenvectors 
(and then the three rotation angles) of $M^{-1}$, can be found by the 
standard procedure of the linear algebra. The eigen-equation of $X$ is 
\be
\label{char}
\left(X-Q_iI\right)\left(\matrix{ V_{1i} \cr V_{2i} \cr V_{3i} }\right)=0
~~~\left(i=1,2,3\right),  
\ee
where $V_{ij}=\left(V_0\right)_{ij}$ and we use $Q_i~(i=1,2,3)$ satisfying
$Q_1>Q_2>Q_3$ to denote the three eigenvalues of $X$. The eigenvectors, 
solution of Eq.\ (\ref{char}), can be expressed in:
\begin{mathletters}
\label{v1}
\begin{eqnarray}
V_{21} &=&	\frac{\left(X_{12}X_{33}-X_{13}X_{23}\right)-Q_1X_{12}}
		{\left(X_{23}^2-X_{33}X_{22}\right)+\left(X_{33}+X_{22}\right)Q_1
		-Q_1^2}V_{11}, \\
V_{31} &=&	\frac{\left(X_{13}X_{22}-X_{12}X_{23}\right)-Q_1X_{13}}
		{\left(X_{23}^2-X_{33}X_{22}\right)+\left(X_{22}+X_{33}\right)Q_1
		-Q_1^2}V_{11}
\end{eqnarray}
\end{mathletters}
and etc. We also know that 
\be
X^{-1}=\frac{1}{\det X}{\rm Adjoint}X.  
\ee
Notice $\det X=1$, the inverse of $X$ is just its adjoint matrix. So  
\begin{eqnarray}
Y_{11}=X_{22}X_{33}-X_{23}^2,~~~~~Y_{22}=X_{11}X_{33}-X_{13}^2,~~~~~
Y_{33}=X_{11}X_{22}-X_{12}^2, \nonumber \\
Y_{12}=X_{13}X_{23}-X_{12}X_{33},~~~~~Y_{13}=X_{12}X_{23}-X_{13}X_{22},~~~~~
Y_{23}=X_{12}X_{13}-X_{11}X_{23}
\end{eqnarray}
and $Y_{ij}=Y_{ji}$. The quadratic terms in Eq.\ (\ref{v1}) are just the
elements of $Y$. By replacing them with $Y_{ij}$ ($i,j=1,2,3$), we have
\begin{mathletters}
\label{v2}
\begin{eqnarray}
V_{21} &=&	\frac{Y_{12}+Q_1X_{12}}{\left(Y_{11}+Q_1X_{11}\right)
		-\left(Q_2^{-1}+Q_3^{-1}\right)}V_{11}, \\
V_{31} &=&	\frac{Y_{13}+Q_1X_{13}}{\left(Y_{11}+Q_1X_{11}\right)
		-\left(Q_2^{-1}+Q_3^{-1}\right)}V_{11}
\end{eqnarray}
\end{mathletters}
Here we have used ${\rm Tr}X=X_{11}+X_{22}+X_{33}=Q_1+Q_2+Q_3$ and 
${\rm det}X=Q_1Q_2Q_3=1$. Thus all the non-diagonal elements of $V_0$ can be 
expressed in a unified form: 
\be
\label{v}
V_{ij}=\frac{Y_{ij}+Q_jX_{ij}}{\left(Y_{jj}+Q_jX_{jj}\right)
	-\hat{Q}_j^{-1}}V_{jj}~~~~~\left(i,j=1,2,3~and~i\neq j\right).
\ee
Here $\hat{Q}_j^{-1}={\rm Tr}Y-Q_j^{-1}$. Considering the normalization 
condition (or unitarity of $V_0$) $V_0V_0^T=V_0^TV_0=I$, all the elements 
can be gotten from Eq.\ (\ref{v}). Then the deduction of the three RH 
angles are direct: $\tan \beta_{23}=\frac{V_{23}}{V_{33}}$, 
$\cos \beta_{13}\sin \beta_{12}=V_{12}$, and $\sin \beta_{13}=V_{13}$.

All the relations obtained, including the masses and the angles, can 
be easily transformed to express the light neutrino parameters in 
$M^{-1}$, $m_D$ and $V_0$. The approach is just to make the following 
exchange $\kappa \leftrightarrow -\eta $, $\xi \leftrightarrow $ $-\xi $ 
and $\theta_{ij}\leftrightarrow \beta_{ij}~\left(1\leq i<j\leq 3\right)$.

\section{neutrino masses and mixings}

The deficit of muon neutrino observed by Super-Kamiokande Collaboration 
and the zenith angle distributions of the data can be explained by 
oscillation between $\nu_{\mu }$ and $\nu_{\tau }$ with the best-fit 
parameters at \cite{atm} 
\be
\left(\sin ^22\theta_{23},\Delta m_{atm}^2\right)
=\left(0.95,5.9\times 10^{-3}{\rm eV}^2\right).
\ee

The $\nu_e-\nu_{\mu }$ explanation to the solar neutrino problem requires 
one set of the parameters (the best fit values) in Table I. corresponding 
to the VO, MSW (including LMA, LOW and SMA) respectively. \cite{bahcall}. 
Here MSW and VO refer to Mikheyev-Smirnov-Wolfenstein matter-enhanced 
oscillations \cite{msw} and vacuum oscillations (so-called just-so 
oscillation) respectively. LMA (SMA) stands for a large (small) mixing 
angle and LOW stands for low probability (or low mass). We assume the 
effective neutrino masses have hierarchical pattern, that is, 
$m_1^{{\rm eff}}\ll m_2^{{\rm eff}}\ll m_3^{{\rm eff}}$. 
So $n_3^2=m_3^{{\rm eff}}\approx \sqrt{\Delta m_{atm.}^2}$ 
and $n_2^2=m_2^{{\rm eff}}\approx \sqrt{\Delta m_{solar}^2}$. 
Little is known about the value of $m_1^{{\rm eff}}$ for which 
we use the parameter $r=\frac{m_2^{{\rm eff}}}{m_1^{{\rm eff}}}\gg 1$ 
to denote. In the framework of three-flavor neutrino oscillations, the big 
hierarchy between $\Delta m_{atm.}^2$ and $\Delta m_{solar}^2$ together 
with the no observation of $\bar{\nu}_e\longrightarrow \bar{\nu}_e$ 
oscillation in the CHOOZ experiment implies that the $\nu_3$-component 
in $\nu_e$ is rather small (even negligible) and the upper limit on the 
value of the $\theta_{13}$ is \cite{up}: 
\be
\sin ^2\theta_{13}\equiv \left| U_{e3}\right| ^2\leq 0.015-0.05.
\ee
We shall therefore set $\theta_{13}=0$. The Dirac masses of neutrino 
are taken at the scale $\mu =10^9{\rm GeV}$ \cite{quark}: 
\be
m_D^{{\rm diag}}\left(\mu \right)
={\rm diag}\left\{ m_u\left(\mu \right),m_c\left(\mu \right),m_t\left(\mu \right)\right\} 
={\rm diag}\left\{ 1.47{\rm MeV},427{\rm MeV},149{\rm GeV}\right\} .
\ee
These are the whole values entering $A$ and $B$. 

\section{Analysis and result}

In this section we start from Eqs.\ (\ref{A:1},\ref{B:1}) to get the RH mass 
hierarchies, $\eta_3$ and $\eta_8$. Then using Eq.\ (\ref{v}), the elements 
(and then the mixing angles) of the RH mixing matrix would be obtained. 
The Majorana masses can be obtained from Eq.\ (\ref{M:1}).

Although we have decoupled the Majorana masses and the RH mixing, the 
expressions of these parameters would be so complicated due to the 
complicated structure of $X$ that it is hard to see explicitly the 
relations of various physical parameters. The hierarchical properties 
of the Dirac and the effective masses of neutrinos make it possible to 
drop the smaller terms in $A$ and $B$. In the following, only the leading 
order terms of $X_{ij}$ ($Y_{ij}$) and $A$ ($B$) will be reserved respectively.

Instead of calculating the RH Majorana parameters by inserting the 
values of these parameters in, we give a more general analysis in two 
cases according to whether $\theta_{12}$ is large (VO, LMA and LOW) or 
small (SMA) and derive the corresponding relations between the masses 
and mixing of the RH neutrino and the other neutrino parameters.

\subsection{Case I: large $\protect\theta_{12}$}
\subsubsection{mass}
In this case, all the elements of $U$ have the same order except that 
$U_{e3}=0$. Reserving the leading order terms in $A$ and $B$, we find 
\begin{mathletters}
\begin{eqnarray}
A &\approx & U_{e2}^2\exp \left(2\xi_3+2\xi_8+2\kappa_3-2\kappa_8\right)
		+U_{\mu 3}^2\exp \left(4\kappa_8-2\xi_3+2\xi_8\right), \\
B &\approx & U_{\tau 1}^2\exp \left(2\kappa_3+2\kappa_8+4\xi_8\right).
\end{eqnarray}
\end{mathletters}
It is easy to see that both $A$ and $B$ are far larger than $3$. Noticing 
that, when $\frac{\Delta m_{atm}^2}{\Delta m_{solar}^2}\leq 10^{8}$, 
we also have $A<B$. Then from Eq.\ (\ref{A<B}) 
one has
\begin{mathletters}
\begin{eqnarray}
Q_1 &=&e^{2\eta_3+2\eta_8}\approx U_{e2}^2
\exp \left(2\kappa_3-2\kappa_8+2\xi_3+2\xi_8\right), \\
Q_2 &=&e^{-2\eta_3+2\eta_8}\approx U_{\mu 3}^2
\exp \left(4\kappa_8-2\xi_3+2\xi_8\right), \\
Q_3 &=&e^{-4\eta_8}\approx \frac{1}{U_{\tau 1}^2}
\exp \left(-2\kappa_3-2\kappa_8-4\xi_8\right).
\end{eqnarray}
\end{mathletters}
Here we have used the relation $U_{e2}^2U_{\mu 3}^2=U_{\tau 1}^2$ which 
is satisfied when $\theta_{13}=0$. We would point out that our results 
would be right as long as $\theta_{13}$ is small enough. Substituting 
the eigenvalues in Eq.\ (\ref{M:1}), we have 
\be
\label{M:2}
M_1\approx \frac{1}{\sin ^2\theta_{12}}\frac{m_{u}^2}{m_2^{{\rm eff}}},~~~~~
M_2\approx \frac{1}{\sin ^2\theta_{23}}\frac{m_{c}^2}{m_3^{{\rm eff}}},~~~~~
M_3\approx \sin ^2\theta_{23}\sin ^2\theta_{12}\frac{m_{t}^2}{m_1^{{\rm eff}}}.  
\ee
The formula are the same as given in Ref. \cite{kuo1}. $M_1$ and $M_2$ 
scale as $1/m_2^{{\rm eff}}$ and $1/m_3^{{\rm eff}}$ respectively while 
$M_3$ scales as $1/m_1^{{\rm eff}}$, which gives scales for the two 
lighter masses, $M_1$ and $M_2$, lower and the heaviest one, $M_3$, 
higher than one would expect when no mixing occurs.

\subsubsection{angles}

Reserving the leading order terms of the numerators and denominators 
in Eq.\ (\ref{v}) respectively, we obtain
\begin{mathletters}
\begin{eqnarray}
V_{21}&\approx &\frac{U_{\mu 2}}{U_{e2}}e^{-2\xi_3}V_{11},~~~~~ 
V_{31}\approx \frac{U_{\tau 2}}{U_{e2}}e^{-\xi_3-3\xi_8}V_{11}, \\
V_{12}&\approx &-\frac{U_{\mu 2}}{U_{e2}}e^{-2\xi_3}V_{22},~~~~~ 
V_{32}\approx -\frac{U_{\mu 1}}{U_{\tau 1}}e^{\xi_3-3\xi_8}V_{22},\\
V_{13}&\approx &\frac{U_{e1}}{U_{\tau 1}}e^{-\xi_3-3\xi_8}V_{33},~~~~~
V_{23}\approx \frac{U_{\mu 1}}{U_{\tau 1}}e^{\xi_3-3\xi_8}V_{33}.
\end{eqnarray}
\end{mathletters}
Exploiting the unitarity of $V_0$, it is appropriate to set $V_{ii}\approx 1$. 
Then the three RH angles are
\begin{mathletters}
\label{angle:1}
\begin{eqnarray}
\beta_{12} &\approx &V_{12}
\approx -\frac{m_{u}}{m_{c}}\cos \theta_{23}\cot \theta_{12},\\
\beta_{13} &\approx &V_{13}
\approx \frac{m_{u}}{m_{t}}\frac{\cot \theta_{12}}{\sin \theta_{23}},\\
\beta_{23} &\approx &V_{23}\approx -\frac{m_{c}}{m_{t}}\cot \theta_{23}.
\end{eqnarray}
\end{mathletters}
All of the RH angles are small and independent of the effective neutrino 
masses. Note that, not like the LH quark mixing where 
$\tan \theta \approx \sqrt{\frac{m_D}{m_{s}}}$ in two-generation case 
\cite{lh}, the RH mixing angles scale linearly with the ratios of the 
Dirac neutrino masses.

\subsubsection{numerical results}

\paragraph{VO}

Inserting the parameters in Eq.\ (\ref{M:2}), we have 
\be
M_1\approx 8.0\times 10^8{\rm GeV},~~~~~
M_2\approx 4.6\times 10^9{\rm GeV},~~~~~
M_3/r\approx 1.5\times 10^{17}{\rm GeV}.
\ee
The mixing angles are easy to obtain from Eq.\ (\ref{angle:1}), 
\be
\beta_{12}\approx -4.6\times 10^{-3},~~~~~
\beta_{13}\approx 3.2\times 10^{-5},~~~~~
\beta_{23}\approx -4.3\times 10^{-3}.
\ee

\paragraph{LMA}

In this case we have nearly the same RH angles as in VO and we find
\be
M_1\approx 1.5\times 10^6{\rm GeV},~~~~~
M_2\approx 4.6\times 10^9{\rm GeV},~~~~~
M_3/r\approx 2.8\times 10^{14}{\rm GeV}.
\ee

\paragraph{LOW}

We now have: 
\be
M_1\approx 1.5\times 10^7{\rm GeV},~~~~~
M_2\approx 4.6\times 10^9{\rm GeV},~~~~~
M_3/r\approx 6.6\times 10^{15}{\rm GeV}
\ee
and 
\be
\beta_{12}\approx -3.3\times 10^{-3},~~~~~
\beta_{13}\approx 2.3\times 10^{-5},~~~~~
\beta_{23}\approx -4.3\times 10^{-3}.
\ee

\subsection{Case II: small $\protect\theta_{12}$ (SMA)}

In this case, $U_{e3}=0$ and $U_{e2}$, $U_{\mu 1}$ and $U_{\tau 1}$ have 
the same order $10^{-2}$ while the other elements of $U$ are of order $1$. 
We have 
\begin{mathletters}
\begin{eqnarray}
A &\approx &U_{e1}^2\exp \left(-2\kappa_3-2\kappa_8+2\xi_3+2\xi_8\right)
+U_{e2}^2\exp \left(2\kappa_3-2\kappa_8+2\xi_3+2\xi_8\right)\approx X_{11}, \\
B &\approx &U_{\tau 2}^2\exp \left(-2\kappa_3+2\kappa_8+4\xi_8\right)
+U_{\tau 1}^2\exp \left(2\kappa_3+2\kappa_8+4\xi_8\right)\approx Y_{33}.
\end{eqnarray}
\end{mathletters}
Again, they satisfy $B>A\gg 3$ and $A^2\gg 4B$. So that
\begin{mathletters}
\begin{eqnarray}
Q_1 &\approx &A\approx f^{-1}U_{e2}^2
\exp \left(2\kappa_3-2\kappa_8+2\xi_3+2\xi_8\right),\\
Q_2 &\approx &\frac{B}{A}\approx U_{\mu 3}^2
\exp \left(e^{4\kappa_3}+4\kappa_8-2\xi_3+2\xi_8\right),\\
Q_3 &\approx &\frac{1}{B}\approx fU_{\tau 1}^2
\exp \left(-2\kappa_3-2\kappa_8-4\xi_8\right).
\end{eqnarray}
\end{mathletters}
Here $f=\frac{r}{r+\cot ^2\theta_{12}}$ and it cannot be omitted 
since $\cot \theta_{12}\gg 1$. Similar with case I, we have 
\be
\label{M:3}
M_1\approx f\frac{1}{\sin ^2\theta_{12}}
\frac{m_{u}^2}{m_2^{{\rm eff}}},~~~~~
M_2\approx \frac{1}{\sin ^2\theta_{23}}
\frac{m_{c}^2}{m_3^{{\rm eff}}},~~~~~
M_3\approx f^{-1}\sin ^2\theta_{23}\sin ^2\theta_{12}
\frac{m_{t}^2}{m_1^{{\rm eff}}}.  
\ee
For the mixing angles, we obtain 
\begin{mathletters}
\begin{eqnarray}
\label{angle:2}
\beta_{12} &\approx &V_{12}\approx -f\frac{m_{u}}{m_{c}}
\cos \theta_{23}\cot \theta_{12},  \\
\beta_{13} &\approx &V_{13}\approx f\frac{m_{u}}{m_{t}}
\frac{\cot \theta_{12}}{\sin \theta_{23}}, \\
\beta_{23} &\approx &V_{23}\approx -\frac{m_{c}}{m_{t}}\cot \theta_{23}.
\end{eqnarray}
\end{mathletters}
Again the factor $f$ appears. Note that the expressions of $M_2$ and 
$\beta_{23}$ are the same as that when $\theta_{12}$ is large. Moreover, 
the SK data suggests strongly that $\theta_{23}\approx \frac{\pi }{4}$. 
So both $M_2$ and $\beta_{23}$ have the same values in all the favored 
regions considered. It is noteworthy that the factor $f$ makes the value 
of $M_3$ remain at a relative low scale for a wide range of $r$ which is 
different with that in Ref.~\cite{kuo1}. When $r\gg \cot ^2\theta_{12}$ 
(then $f\approx 1$), we have the same expressions of the RH masses and 
the mixing angles no matter whether $\theta_{12}$ is large or not.

Substituting the values of the parameters in, from Eq.\ (\ref{M:3}) we have
\be
M_1\approx 4.7\times 10^8f{\rm GeV},~~~~~ 
M_2\approx 4.6\times 10^9{\rm GeV},~~~~~
M_3\approx 3.0\times 10^{12}\frac{r}{f}{\rm GeV}.
\ee
and from Eq.\ (\ref{angle:2}),
\be
\beta_{12}\approx -7.0\times 10^{-2}f,~~~~~
\beta_{13}\approx 4.9\times 10^{-4}f,~~~~~
\beta_{23}\approx -4.3\times 10^{-3}.
\ee
Here, with the value of $\theta_{12}$ substituted in, 
$f\approx \frac{r}{r+6.6\times 10^2}$.

Comparisons with the exact numerical results are given in Tables II-IV 
and from which we can see they fit well. In calculation we take 
$m_D^{{\rm diag}}\left(\mu \right)$ at $\mu =10^9{\rm GeV}$. Note that, 
although the up quark masses are running with $\mu$, the dirac mass 
hierarchies, $\eta_3$ and $\eta_8$, are almost fixed when $\mu $ varies. 
We find they satisfy the following approximate relation 
\be
\frac{m_{u}\left(\mu \right)m_{t}\left(\mu \right)}{m_{c}^2\left(\mu \right)}\approx 1.
\ee
So the deviation is mainly resulted from $F$
($=\frac{m_{t}^2}{m_3^{{\rm eff}}}e^{4\kappa_8-4\xi_8}$) 
when $m_D^{{\rm diag}}\left(\mu \right)$ at different scale is taken.


\section{Summary and Discussion}

In this paper, we introduce a parameterization which transforms all the 
involving masses in the seesaw formula to the mass ratios. Then by taking 
the traces of $X$ and its inverse, we derive the equations of the Majorana 
mass ratios, $\eta_3$ and $\eta_8$. The solutions to these equations are 
obtained under some conditions and the elements of $V_0$ are expressed in 
a unified form. Assuming quark-lepton symmetry and hierarchical effective 
neutrino masses, rather simple relations among the various neutrino parameters 
entering the seesaw formula are deduced. Finally, setting the Dirac neutrino 
masses to be equal to the up quark masses, we present the numerical results 
in the favored regions of the solar and atmospheric neutrino experiments.

Now let us give a combined analysis of the results obtained and list our
main points as follows:

\begin{itemize}
\item $M_2$($\approx 4.6\times 10^9{\rm GeV}$) and so the product of $M_1$ 
and $M_3$ are nearly independent of $\theta_{12}$.

\item The three RH neutrino masses are hierarchical and $\frac{M_3}{M_2}
\left(\propto \frac{m_3^{{\rm eff}}}{m_1^{{\rm eff}}}\right)
\gg \frac{M_2}{M_1}
\left(\propto \frac{m_1^{{\rm eff}}}{m_2^{{\rm eff}}}\right)$.

\item $\beta_{23}$ ($\approx -4.3\times 10^{-3}$ ) and 
$\beta_{12}/\beta_{13}
\approx -\frac{1}{2}\frac{m_{t}}{m_{c}}\sin 2\theta_{23}
\approx -\frac{1}{2}\frac{m_{t}}{m_{c}}$ are also independent of 
$\theta_{12}$. Moreover, the RH mixing angles satisfy the following condition
\be
\frac{\beta_{12}\beta_{23}}{\beta_{13}}
\approx \cos ^2\theta_{23}\approx \frac{1}{2}
\ee
which is independent of not only $\theta_{12}$ and the effective neutrino 
masses but also the Dirac masses of neutrinos. It is interesting to notice 
that the $\left(13\right)$ elements ($U_{e3}$, $V_{13}$ and $U_{us}$) 
determined by the third mixing angles of the three corresponding mixing 
matrices are all small. It is also noteworthy that the third mixing angles 
in both the CKM matrix of quarks and the RH mixing matrix are of orders of 
the products of the other two angles respectively. In the former, we have 
$\left| \frac{U_{us}U_{ub}}{U_{cb}}\right| \approx 
\left(\rho ^2+\eta ^2\right)^{-\frac{1}{2}}$. Here, $\rho$ and $\eta$ are 
smaller than one \cite{fritzsch}.

\item Numerically, the lightest right-handed neutrino mass can lie between 
$10^6{\rm GeV}$ and $10^8{\rm GeV}$ while the heaviest right-handed neutrino 
mass range from about $10^{12}{\rm GeV}$ to far larger than $10^{17}{\rm GeV}$. 

\item Numerically, all the three RH angles are small although they may
contain the contribution from the diagonalization of $M^{-1}$. The absolute 
values of $\beta_{12}$ and $\beta_{13}$ are about $10^{-3}\sim 10^{-2}$ and 
$10^{-6}\sim 10^{-4}$ respectively.

\item SMA solution seems especially attractive in the sense that 
$M_3\sim 10^{15}{\rm GeV}$ for a wide range of $r$ due to the factor $f$ 
while $M_3$'s for the other three regions (VO, LMA and LOW) increase rapidly 
with $r$ and become too large to be viable. Especially, for the VO solution 
to the solar neutrino problem, both the two mass squared differences splittings 
(of the order $10^{-3}{\rm eV}^2$ and $10^{-11}{\rm eV}^2$ respectively) and 
the scale of the heaviest RH neutrino mass $M_3$ ($\gg 10^{17}{\rm GeV}$) make 
it look very unnatural.
\end{itemize}

In this work, we have set $\theta_{13}=0$. Although the small $\theta_{13}$ 
has little effect on the oscillation solution to the solar and the atmospheric 
neutrino deficits, it may become important in the seesaw mechanism especially 
in the SMA region where $\theta_{13}$ is comparable with $\theta_{12}$. It may 
lead to large RH mixing angles owing to the contribution from the diagonalization 
of $M^{-1}$ as well as degenerate masses. This can also be seen from that 
the coefficient of $U_{e3}$ in $A$ are much larger than that of the other 
elements of $U$. We point out that the method is even valid in such case 
while more skills may be needed. We will discuss it in more details in later paper.

\acknowledgements

The authors would like to express our sincere thanks to Professor T.~K.~Kuo 
for pointing out this problem to G.~Cheng during his visit at Purdue University 
from January to April, 2000 and provoke our interesting in it. We are also 
indebted to him for his warmly help in the research progressing and kindness 
by giving us his papers being prepared. We are grateful to Dr.~Du~Taijiao and 
Dr.~Tu~Tao for useful discussions. The authors are also grateful to the 
convenience provided by Chinese High Performance Computing Center at Hefei. 
The authors are supported in part by the National Science Foundation in China 
grant no.19875047.

\addtolength{\baselineskip}{-.3\baselineskip}

\begin{table}[tbp]
{\small Table I: $\nu_{e}\rightarrow \nu_{\mu }$ solutions to the solar
neutrino problem. Here MSW and VO refer to Mikheyev-Smirnov-Wolfenstein
matter-enhanced oscillations \cite{msw} and vacuum oscillations (so-called
just-so oscillation) respectively. LMA (SMA) stands for a large (small)
mixing angle and LOW stands for low probability (or low mass).}
\par
\begin{center}
\begin{tabular}{ccc}
\hline\hline
Solution & $\Delta m_{solar}^2\left({\rm eV}^2\right)$ 
& $\sin ^22\theta_{12}$ \\ \hline
VO & $6.5\times 10^{-11}$ & $0.75$ \\ 
MSW(LMA) & $1.8\times 10^{-5}$ & $0.76$ \\ 
MSW(LOW) & $7.9\times 10^{-8}$ & $0.96$ \\ 
MSW(SMA) & $5.4\times 10^{-6}$ & $6.0\times 10^{-3}$ \\ \hline
\end{tabular}
\end{center}
\end{table}

\begin{table}[tbp]
{\small Table II: Exact numerical and approximate results when $r=10^1$.
In each cell we listed the numerical and approximate results above and below respectively. By solving the eigen-equation of $X$ we obtain the eigenvalue(s) that larger than one and the corresponding eigenvector(s). The reciprocal value(s) of the other eigenvalue(s) and the corresponding eigenvector(s) are obtained by solving the eigen-equation of $Y$. Substituting these eigenvalues in Eq.(24) we get the three masses in Majorana sector (see the text for details). }
\par
\begin{center}
\begin{tabular}[t]{ccccccc}
\hline
$r=10^1$ & $M_1\left({\rm GeV} \right)$ 
& $M_2\left({\rm GeV} \right)$ & $M_3\left({\rm GeV} \right)$ 
& $\beta_{12}$ & $\beta_{13}$ & $\beta_{23}$ \\ \hline
VO & $
\begin{array}{c}
6.2\times 10^8 \\ 8.0\times 10^8
\end{array}
$ & $
\begin{array}{c}
4.6\times 10^9 \\ 4.6\times 10^9
\end{array}
$ & $
\begin{array}{c}
1.9\times 10^{18} \\ 1.5\times 10^{18}
\end{array}
$ & $
\begin{array}{c}
-3.7\times 10^{-3} \\ -4.6\times 10^{-3}
\end{array}
$ & $
\begin{array}{c}
2.2\times 10^{-5} \\ 3.2\times 10^{-5}
\end{array}
$ & $
\begin{array}{c}
-4.3\times 10^{-3} \\ -4.3\times 10^{-3}
\end{array}
$ \\ \hline
LMA & $
\begin{array}{c}
1.2\times 10^6 \\ 1.5\times 10^6
\end{array}
$ & $
\begin{array}{c}
4.5\times 10^9 \\ 4.6\times 10^9
\end{array}
$ & $
\begin{array}{c}
3.7\times 10^{15} \\ 2.8\times 10^{15}
\end{array}
$ & $
\begin{array}{c}
-3.2\times 10^{-3} \\ -4.6\times 10^{-3}
\end{array}
$ & $
\begin{array}{c}
2.2\times 10^{-5} \\ 3.2\times 10^{-5}
\end{array}
$ & $
\begin{array}{c}
-4.1\times 10^{-3} \\ -4.3\times 10^{-3}
\end{array}
$ \\ \hline
LOW & $
\begin{array}{c}
1.3\times 10^7 \\ 1.5\times 10^7
\end{array}
$ & $
\begin{array}{c}
4.6\times 10^9 \\ 4.6\times 10^9
\end{array}
$ & $
\begin{array}{c}
7.6\times 10^{16} \\ 6.6\times 10^{16}
\end{array}
$ & $
\begin{array}{c}
-2.6\times 10^{-3} \\ -3.3\times 10^{-3}
\end{array}
$ & $
\begin{array}{c}
1.8\times 10^{-5} \\ 2.3\times 10^{-5}
\end{array}
$ & $
\begin{array}{c}
-4.3\times 10^{-3} \\ -4.3\times 10^{-3}
\end{array}
$ \\ \hline
SMA & $
\begin{array}{c}
7.0\times 10^6 \\ 7.0\times 10^6
\end{array}
$ & $
\begin{array}{c}
4.4\times 10^9 \\ 4.6\times 10^9
\end{array}
$ & $
\begin{array}{c}
2.1\times 10^{15} \\ 2.0\times 10^{15}
\end{array}
$ & $
\begin{array}{c}
-9.3\times 10^{-4} \\ -1.0\times 10^{-3}
\end{array}
$ & $
\begin{array}{c}
6.2\times 10^{-6} \\ 7.2\times 10^{-6}
\end{array}
$ & $
\begin{array}{c}
-4.0\times 10^{-3} \\ -4.3\times 10^{-3}
\end{array}
$ \\ \hline
\end{tabular}
\end{center}
\end{table}

\begin{table}[tbp]
{\small Table III: Same as in table I but for $r=10^2$. }
\par
\begin{center}
\begin{tabular}[t]{ccccccc}
\hline
$r=10^2$ & $M_1\left({\rm GeV}\right)$ 
& $M_2\left({\rm GeV}\right)$ & $M_3\left({\rm GeV}\right)$ 
& $\beta_{12}$ & $\beta_{13}$ & $\beta_{23}$ \\ \hline
VO & $
\begin{array}{c}
7.7\times 10^8 \\ 8.0\times 10^8
\end{array}
$ & $
\begin{array}{c}
4.6\times 10^9 \\ 4.6\times 10^9
\end{array}
$ & $
\begin{array}{c}
1.5\times 10^{19} \\ 1.5\times 10^{19}
\end{array}
$ & $
\begin{array}{c}
-5.3\times 10^{-3} \\ -4.6\times 10^{-3}
\end{array}
$ & $
\begin{array}{c}
3.1\times 10^{-5} \\ 3.2\times 10^{-5}
\end{array}
$ & $
\begin{array}{c}
-4.3\times 10^{-3} \\ -4.3\times 10^{-3}
\end{array}
$ \\ \hline
LMA & $
\begin{array}{c}
1.5\times 10^6 \\ 1.5\times 10^6
\end{array}
$ & $
\begin{array}{c}
4.6\times 10^9 \\ 4.6\times 10^9
\end{array}
$ & $
\begin{array}{c}
2.9\times 10^{16} \\ 2.8\times 10^{16}
\end{array}
$ & $
\begin{array}{c}
-4.4\times 10^{-3} \\ -4.6\times 10^{-3}
\end{array}
$ & $
\begin{array}{c}
3.1\times 10^{-5} \\ 3.2\times 10^{-5}
\end{array}
$ & $
\begin{array}{c}
-4.3\times 10^{-3} \\ -4.3\times 10^{-3}
\end{array}
$ \\ \hline
LOW & $
\begin{array}{c}
1.4\times 10^7 \\ 1.5\times 10^7
\end{array}
$ & $
\begin{array}{c}
4.6\times 10^9 \\ 4.6\times 10^9
\end{array}
$ & $
\begin{array}{c}
6.7\times 10^{17} \\ 6.6\times 10^{17}
\end{array}
$ & $
\begin{array}{c}
-3.2\times 10^{-3} \\ -3.3\times 10^{-3}
\end{array}
$ & $
\begin{array}{c}
2.3\times 10^{-5} \\ 2.3\times 10^{-5}
\end{array}
$ & $
\begin{array}{c}
-4.3\times 10^{-3} \\ -4.3\times 10^{-3}
\end{array}
$ \\ \hline
SMA & $
\begin{array}{c}
6.1\times 10^7 \\ 6.1\times 10^7
\end{array}
$ & $
\begin{array}{c}
4.4\times 10^9 \\ 4.6\times 10^9
\end{array}
$ & $
\begin{array}{c}
2.4\times 10^{15} \\ 2.3\times 10^{15}
\end{array}
$ & $
\begin{array}{c}
-9.1\times 10^{-3} \\ -9.1\times 10^{-3}
\end{array}
$ & $
\begin{array}{c}
6.0\times 10^{-5} \\ 6.4\times 10^{-5}
\end{array}
$ & $
\begin{array}{c}
-4.0\times 10^{-3} \\ -4.3\times 10^{-3}
\end{array}
$ \\ \hline
\end{tabular}
\end{center}
\end{table}

\begin{table}[tbp]
{\small Table IV: Same as in table I but for $r=10^3$. }
\par
\begin{center}
\begin{tabular}[t]{ccccccc}
\hline
$r=10^3$ & $M_1\left({\rm GeV}\right)$ 
& $M_2\left({\rm GeV}\right)$ & $M_3\left({\rm GeV}\right)$ 
& $\beta_{12}$ & $\beta_{13}$ & $\beta_{23}$ \\ \hline
VO & $
\begin{array}{c}
7.9\times 10^8 \\ 8.0\times 10^8
\end{array}
$ & $
\begin{array}{c}
4.6\times 10^9 \\ 4.6\times 10^9
\end{array}
$ & $
\begin{array}{c}
1.5\times 10^{20} \\ 1.5\times 10^{20}
\end{array}
$ & $
\begin{array}{c}
-5.5\times 10^{-3} \\ -4.6\times 10^{-3}
\end{array}
$ & $
\begin{array}{c}
3.2\times 10^{-5} \\ 3.2\times 10^{-5}
\end{array}
$ & $
\begin{array}{c}
-4.3\times 10^{-3} \\ -4.3\times 10^{-3}
\end{array}
$ \\ \hline
LMA & $
\begin{array}{c}
1.5\times 10^6 \\ 1.5\times 10^6
\end{array}
$ & $
\begin{array}{c}
4.5\times 10^9 \\ 4.6\times 10^9
\end{array}
$ & $
\begin{array}{c}
2.8\times 10^{17} \\ 2.8\times 10^{17}
\end{array}
$ & $
\begin{array}{c}
-4.6\times 10^{-3} \\ -4.6\times 10^{-3}
\end{array}
$ & $
\begin{array}{c}
3.2\times 10^{-5} \\ 3.2\times 10^{-5}
\end{array}
$ & $
\begin{array}{c}
-4.3\times 10^{-3} \\ -4.3\times 10^{-3}
\end{array}
$ \\ \hline
LOW & $
\begin{array}{c}
1.5\times 10^7 \\ 1.5\times 10^7
\end{array}
$ & $
\begin{array}{c}
4.6\times 10^9 \\ 4.6\times 10^9
\end{array}
$ & $
\begin{array}{c}
6.6\times 10^{18} \\ 6.6\times 10^{18}
\end{array}
$ & $
\begin{array}{c}
-3.3\times 10^{-3} \\ -3.3\times 10^{-3}
\end{array}
$ & $
\begin{array}{c}
2.3\times 10^{-5} \\ 2.3\times 10^{-5}
\end{array}
$ & $
\begin{array}{c}
-4.3\times 10^{-3} \\ -4.3\times 10^{-3}
\end{array}
$ \\ \hline
SMA & $
\begin{array}{c}
2.8\times 10^8 \\ 2.8\times 10^8
\end{array}
$ & $
\begin{array}{c}
4.5\times 10^9 \\ 4.6\times 10^9
\end{array}
$ & $
\begin{array}{c}
5.1\times 10^{15} \\ 5.0\times 10^{15}
\end{array}
$ & $
\begin{array}{c}
-4.4\times 10^{-2} \\ -4.2\times 10^{-2}
\end{array}
$ & $
\begin{array}{c}
2.9\times 10^{-4} \\ 2.9\times 10^{-4}
\end{array}
$ & $
\begin{array}{c}
-4.1\times 10^{-3} \\ -4.3\times 10^{-3}
\end{array}
$ \\ \hline
\end{tabular}
\end{center}
\end{table}

\end{document}